\documentclass[runningheads]{llncs}

\usepackage{graphicx}
\usepackage[x11names]{xcolor}
\usepackage[hidelinks]{hyperref}
\usepackage{todonotes}
\usepackage{listings}
\usepackage{subcaption}
\usepackage{cite}
\usepackage{enumitem}
\usepackage[pdf]{graphviz}
\usepackage{moresize}
\usepackage{booktabs}

\usepackage{xpatch}
\makeatletter
\newcommand*{\addFileDependency}[1]{
  \typeout{(#1)}
  \@addtofilelist{#1}
  \IfFileExists{#1}{}{\typeout{No file #1.}}
}
\makeatother
\xpretocmd{\digraph}{\addFileDependency{#2.dot}}{}{}

\graphicspath{ {./figures/} }

\usepackage{amsfonts}
\DeclareMathSymbol{\mhyphen}{\mathord}{AMSa}{"39}

\lstdefinelanguage{scala}{
  alsoletter={@},
  morekeywords={abstract, case, catch, choose, class, def, do, else, extends, final, finally, for, if, implicit, import, match, new, null, object, let, in, be, lazy,holds,@erasable,
override, package, private, protected, requires, ensures, decreases, return, sealed, super, then, this, throw, trait, try, type, val, var, while, yield, domain, template, assume, fun,
postcondition, precondition,invariant, constraint, assert, each, _, return, @generator, ensure, require, ensuring, assuming, otherwise, asserting, enum}
  sensitive=true,
  morecomment=[l]{//},
  morecomment=[s]{/*}{*/},
  morestring=[b]",
  literate={{=>}{$\Rightarrow$\,}2%
            {->}{$\rightarrow$\,}2%
            {<=}{$\le$\,}2%
            {>=}{$\ge$\,}2%
            {<-}{$\leftarrow$\,}2%
            }
}

\newcommand{\codestyle}{\small\sffamily}

\newcommand{\commentstyle}{\small\color{teal}\textsl}

\newcommand{\asncommentstyle}{\small\color{blue}}

\lstdefinelanguage{asn}{
  keywords={
  OF, SEQUENCE, SIZE, CHOICE, INTEGER
  },
  sensitive=true,
  morecomment=[l]{//},
  morecomment=[s]{/*}{*/},
  morecomment=[s]{[}{]},
  morestring=[b]",
  columns=flexible,
  commentstyle=\asncommentstyle,
  literate={{<}{$\langle$}1%
            {>}{$\rangle$}1%
            {-}{$\mhyphen$}1%
            }
}

\begin{document}

\sloppy

\title{Formally Verifiable Generated ASN.1/ACN Encoders and Decoders: A Case Study}

\titlerunning{Formally Verifiable Generated ASN.1/ACN Encoders and Decoders}

\author{Mario Bucev\inst{1} \and
Samuel Chassot\inst{1}\orcidID{0009-0000-9751-9252} \and
Simon Felix\inst{2}\orcidID{0000-0002-3979-128X} \and
Filip Schramka\inst{2} \and
Viktor Kun\v{c}ak\inst{1}\orcidID{0000-0001-7044-9522}
}

\authorrunning{M.~Bucev, S.~Chassot, S.~Felix, F.~Schramka, V.~Kun\v{c}ak}

\institute{EPFL IC LARA, Lausanne, Switzerland\\
\email{mario.bucev@epfl.ch samuel.chassot@epfl.ch viktor.kuncak@epfl.ch}\\
\and
Ateleris GmbH, Brugg, Switzerland\\
\email{simon.felix@ateleris.ch filip.schramka@ateleris.ch}}

\maketitle

\begin{abstract}
We propose a verified executable Scala backend for ASN1SCC, a compiler for ASN.1/ACN. ASN.1 is a language for describing data structures widely employed in ground and space telecommunications. ACN can be used along ASN.1 to describe complex binary formats and legacy protocols. To avoid error-prone and time-consuming manual writing of serializers, we show how to port an ASN.1/ACN code generator to generate Scala code. We then enhance the generator to emit not only the executable code but also strong enough preconditions, postconditions, and lemmas for inductive proofs. This allowed us to verify the resulting generated annotated code using Stainless, a program verifier for Scala. The properties we prove include the absence of runtime errors, such as out-of-bound accesses or divisions by zero. For the base library, we also prove the invertibility of the decoding and encoding functions, showing that decoding yields the encoded value back. Furthermore, our system automatically inserts invertibility proofs for arbitrary records in the generated code, proving over 300'000 verification conditions. We establish key steps towards such proofs for sums and arrays as well.
\keywords{formal verification \and serialization \and Scala \and Stainless}
\end{abstract}

\section{Introduction}

Values of structured data types are a key for programming above the level of assembly. To transmit structured data across a communication channel, programs need to serialize data into bits on one side and deserialize it on the other side.
Writing serialization and deserialization of protocols requires great expertise and is error-prone. Buffer overflows during deserialization may result in security exploits\cite{cwetop25, postquantumbroken2024}. Data corruption and loss may also arise due to hard-to-test implementation errors in serialization or deserialization.
Automated code generation mitigates these issues by letting protocol designers define the messages rather than writing the encoders and decoders by hand. However, bugs may still arise in such an approach due to errors in the code generator or bugs in the used base libraries. We aim to prevent such errors using formal verification.

We use ASN.1 \cite{asn1}, an interface description language (IDL)
that specifies protocol messages, as well as 
their serialization and deserialization.
ASN.1 describes messages using a declarative specification similar in purpose to algebraic data types, JSON or XML schemas. ASN.1 is widely used in many applications, for example, to define the format of HTTPS certificates \cite{x509https} or 5G network packets.

In contrast to other general message formats, such as Protocol Buffers \cite{protobuf} or Apache Thrift \cite{slee2007thrift}, ASN.1 supports the serialization of messages in multiple concrete representations. While this design introduces additional complexity, it also means that users can optimize the wire format for each use case. Out of all ASN.1 wire formats, ACN \cite{mamais_asn1_2012} is the most flexible one because it gives users granular control of most aspects of serialization. This flexibility means that most existing and legacy protocols can be described with ASN.1/ACN.

In this paper, we present the development of an automated generator for verifiable serializers and deserializers for messages conforming to a given ASN.1/ACN specification. The generator emits code in Scala along with inductive specifications in such a way that the generated code automatically verifies using the Stainless verifier \cite{10.1145/3360592}. Stainless is non-interactive: if it times out during verification, the programmer needs to update the source code with additional assertions, preconditions, and postconditions and re-run the verification.

We can consider several levels of verification, depending on the level of properties we expect from code given to the Stainless verifier:
\newcommand{\Level}[1]{{\textbf{Level #1.}}}
\begin{enumerate}
    \item[]\Level{1} The generated subset of Scala corresponds to what Stainless accepts in terms of syntactic constructs and types. This implies the absence of null dereferences, which are ruled out by construction, thanks to use of case class constructors and required declaration-time initialization.
    \item[]\Level{2} Stainless proves all automatically generated verification conditions for generated Scala programs. These conditions guarantee function termination and the absence of run-time errors such as pattern matching failures, array out-of-bounds accesses, division by zero, or unsafe casts. As the verification processes each function individually, in practice, this requires adding appropriate preconditions and postconditions as part of generated functions.
    \item[]\Level{3} Verifying additional partial specification properties. A particularly interesting property verified at this level is the exact number of bits encoded and decoded by each function.
    \item[]\Level{4} Verification of key functional correctness characterization properties, such as the fact that decoding an encoded value (with an appropriate position of the decoder) recovers the original value.
\end{enumerate}
In our work, we first developed an executable Scala backend and a suite of unit tests. The test suite checks whether the generated Scala code produces bit-wise identical results as other ASN.1 serializers. To achieve the first verification level, we needed to modify the code generator to ensure that the initialization of data structures is compositional.
Due to limitations regarding nested mutable types in Stainless, we moved to immutable vectors instead of arrays to make verification more feasible. For level 2, we enhanced the code generator to emit sufficient preconditions and postconditions that ensure the absence of run-time errors.
To ensure verification succeeds in a reasonable amount of time, we modified the code generator to be more compositional in the generated code. We also automatically generate the declarations and uses of lemmas for inductive properties needed to establish the absence of errors. Moving to level 3 required stronger postconditions and additional lemmas. These lemmas are parametric in the data types that are being generated, so they also need to be automatically generated, as opposed to being developed once and for all.

We completed level 3 for generated code, ensuring no run-time errors and that the expected amount of data is written.
We also made significant progress towards level 4: our base library has full specifications of invertibility that are proven by Stainless. Furthermore, we generate verifiable code with specifications that preserve the invertibility properties for records that are encoded serially. This development required additional lemmas about encoding in array buffers. We believe that extending the invertibility proof for the remaining recursive cases is feasible. During the verification effort, we also identified and fixed errors in the existing code generator.

\subsection{Contributions}

This paper makes the following contributions:
\begin{itemize}
    \item We extend the ASN1SCC compiler with a Scala backend alongside a runtime library for encoding and decoding primitive ASN.1/ACN constructs and provide a testing framework to test the interoperability of the newly developed Scala backend with the existing C backend.
    \item We discover and address bugs in existing and new parts of the code generator and library, in part thanks to formal verification (Section~\ref{sec:identified-bugs}).
    \item We prove the runtime safety of the serialization for primitives and the safety of the generated code. Furthermore, we prove strong properties such as invertibility for all library functions except for floating point and string-related operations.  \item We evaluate the automation of our approach by generating encoding and decoding functions for the PUS-C format that is of practical interest in aerospace applications. We prove invertibility for records in the generated code. For all of the generated code, Stainless obtains our proofs automatically because the generated code contains sufficient assertions and other proof hints for Stainless to succeed
    (see Table~\ref{tab:vcs}).
\end{itemize}

Our code generator that produces verifiable Stainless code is available as part of the ASN1SCC generator written in $F^\#$:
\begin{center}
    \url{https://github.com/maxime-esa/asn1scc}
\end{center}
Generated Scala serializers and desearializers for packets defined by the PUS-C standard \cite{pusc} are available at:
\begin{center}
    \url{https://github.com/epfl-lara/fovcom}
\end{center}

\subsection{Related Work}
\label{sec:related}

Narcissus project~\cite{narcissus} developed flexible and trustworthy combinators for parsing and unparsing within the Coq proof assistant, showing that the approach works well enough to replace packet processors for a full Internet protocol stack in the Mirage operating system. Our approach is based on a code generator and starts from ASN.1 and ACN as existing definition languages. We consider it a promising yet challenging future work to develop verified invertible combinators that could express ACN constructs.

The authors of EverParse~\cite{everparse} present a formally verified library and framework for binary formats in F$^*$. This framework supports proofs for non-malleability (unique representation of values), safety and inversion of encoding. EverParse can then emit high-performance C code from the formally proven F$^*$.

Based on EverParse and related to ASN.1 is ASN1*~\cite{ni_asn1_2023}, which supports the Distinguished Encoding Rule (DER) format. This encoding ensures non-malleability, a characteristic that is, for instance, particularly desired in cryptography. The ASN1* authors formalized the DER semantics in F$^*$ within the EverParse framework and proved its non-malleability. The parser combinators are non-malleable and correct by construction.
In contrast to DER, our work supports serialization in the highly flexible ACN \cite{mamais_asn1_2012} format, more intended towards legacy protocols or complex binary format where non-malleability is not required.

Promiwag \cite{mondet2011generating} is a library to generate protocol deserialization code. The authors prove that the generated code will terminate and is free of out-of-bounds memory accesses. Our work proves stronger invariants for the generated code, for example, invertibility, and also covers serialization.
\section{Use of ASN.1 in Space Missions}

Reliable communication is crucial for spacecrafts due to their long missions, large distances, and high costs. In the past, projects relied on human-readable protocol specifications and manually written protocol code on the spacecraft and in the ground software. This made it challenging to iterate a protocol quickly and reliably.

In addition, commercial or military space applications are becoming increasingly attractive targets for attackers. In recent years, the resilience of satellites and ground stations has gained considerable attention from operators and space agencies. According to the CWE Top 25 list \cite{cwetop25}, bugs in the deserialization of untrusted data (CWE-502) are among the most common security vulnerabilities.

To reduce these risks, missions increasingly rely on automated code generators such as ASN1SCC. The European Space Agency (ESA) has announced its cybersecurity strategy \cite{esaCyber}, which also encourages the use of code generators as part of the secure-as-built principle. ESA has produced the PUS-C standard \cite{pusc}, which defines the communication protocol for all ESA missions. The PUS-C standard was later formally specified in ASN.1/ACN \cite{pusclib}. Multiple ESA missions already use automatically generated code, for example, CHEOPS or PROBA-3 \cite{thanassis2018taste}, produced by the ASN1SCC code generator from the PUS-C ASN.1/ACN specification.

The ASN.1 encoding schemes such as PER, uPER or BER control the binary format of the messages. ASN.1 Control Notation\cite{mamais_asn1_2012} (ACN) is an encoding scheme that allows developers to specify the binary layout of data structures, for instance, the determinant of \lstinline{CHOICE} (a sum type), the length of \lstinline{SEQUENCE OF} (an ordered collection), integer bit size, or alignment of fields. This detailed control means that legacy formats with non-uniform formatting rules can be described, for example, PUS-C \cite{pusclib}.

As an example, consider the \texttt{TC[2, 7]} telecommand, which can be defined with ASN.1 as follows:
\begin{lstlisting}[language=asn]
TC-2-7-DistrPhysicalDevCmds ::= SEQUENCE {
  physicalDevCmds SEQUENCE (SIZE(1 .. 63)) OF PhysicalDevCmd
}
PhysicalDevCmd ::= SEQUENCE {protoData ProtoData, cmdData CmdData}
CmdData         ::= CHOICE {dev1 INTEGER (0 .. 255)}
ProtoData       ::= CHOICE {dev1 INTEGER (0 .. 255)}
\end{lstlisting}

\lstinline{TC-2-7-DistrPhysicalDevCmds} is a record containing an array of \lstinline{Physical}\-\lstinline{DevCmd}, whose size ranges between 1 and 63.
\lstinline{CmdData} and \lstinline{ProtoData} are both a sum type with only one variant, an integer ranging from 0 to 255 (note that these are mission-specific and are expected to be tailored).

However, the \texttt{TC[2, 4]} telecommand mandates a certain binary format. In particular, both \lstinline{CmdData} and \lstinline{ProtoData} have their determinant specified by a ``physical device ID'' residing outside of their definition (i.e. the determinant is not embedded within the \lstinline{CHOICE}).
This can be achieved by parameterizing both \lstinline{CHOICE}s and inserting an \emph{ACN field} named \lstinline{physicalDev-ID} within \lstinline{PhysicalDevCmd}:
\begin{lstlisting}[language=asn]
TC-2-7-DistrPhysicalDevCmds [] {n PUSC-UINT32 [], physicalDevCmds [size n]}
PhysicalDevCmd [] {
  physicalDev-ID PhysicalDev-ID [],
  protoData<physicalDev-ID> [], cmdData<physicalDev-ID> []
}
CmdData<PhysicalDev-ID: device> [determinant device] {
  dev1 [/*...*/]
}
ProtoData<PhysicalDev-ID: device> [determinant device] {
  dev1 [/*...*/]
}
\end{lstlisting}

The ACN specification resides in a corresponding \lstinline{.acn} file.
Each definition, field, or variant may have an encoding property specified within brackets; if none are desired, these must be empty.
ACN allows inserting additional fields, for instance, the \lstinline{n} in \lstinline{TC-2-7-DistrPhysicalDevCmds} and \lstinline{physicalDev-ID} in \lstinline{PhysicalDevCmd}. The former is encoding the size of \lstinline{physicalDevCmds} while the latter is encoding the determinant of both \lstinline{protoData} and \lstinline{cmdData}.
Setting the determinant of a \lstinline{CHOICE} is done by parameterizing its definition (by adding a parameter \lstinline{PhysicalDev-ID: device}) and indicating it as an encoding property (done with \lstinline{[determinant device]}).
Finally, each reference to such parameterized \lstinline{CHOICE} must be "instantiated"; in the above example, this corresponds to \lstinline{protoData<physicalDev-ID>} and similarly for \lstinline{cmdData<physicalDev-ID>}.

\section{The Stainless Verification Framework}

We use the Stainless program verifier\footnote{\url{https://github.com/epfl-lara/stainless}} to verify generated encoders and decoders.
Stainless is an open-source deductive verifier for the Scala programming language. The verifier runs the Scala
compiler and interprets certain annotations and function calls (such as {\tt require}, {\tt ensuring}) as specifications. The foundation of Stainless is System FR, a dependent-type extension of System F with refinement types and indexed recursive data type definitions \cite{10.1145/3360592}.
Stainless transforms syntax trees obtained from the Scala compiler into a simpler form, eliminating non-aliased imperative state~\cite{Blanc:230242,BlancETAL13VerificationTranslationRecursiveFunctions} and encoding non-disjoint types \cite[Chapter 5]{DBLP:phd/ch/Voirol19}, then uses a type checker for System FR \cite{10.1145/3360592} to generate proof obligations that contain calls to higher-order and recursive functions, which it passes to its subsystem named Inox~\footnote{\url{https://github.com/epfl-lara/inox}}. Inox unfolds recursive functions~\cite{SuterETAL11SatisfiabilityModuloRecursivePrograms} and resolves higher-order function calls \cite{VoirolETAL15CounterExampleCompleteVerificationHigherOrderFunctions} until it generates quantifier-free first-order queries for Z3~\cite{z3}, CVC4~\cite{cvc4}, cvc5 \cite{DBLP:conf/tacas/BarbosaBBKLMMMN22} and Princess~\cite{princess08} solvers. In each unrolling, Inox alternates checks for validity and counterexamples of verification conditions, allowing Stainless to provide useful feedback to users. It also uses formula simplification, normalization and caching to improve verification efficiency~\cite{GuilloudETAL23FormulaNormalizationsVerification}.

Stainless programs are valid Scala programs and can thus be compiled to run on the JVM, to native code via Scala Native for LLVM, and to JavaScript via Scala.js. Furthermore, Stainless can transpile a subset of programs to simple C code suitable for embedded applications \cite{HamzaETAL22NFM}.
In recent years, Stainless was used to verify several case studies, including key parts of a flash file system for the X-ray spectrometer of the Solar Orbiter satellite
\cite{HamzaETAL22NFM}, a hash table implementation from Scala standard collection library \cite{ChassotKuncak2024VerifiedLongmap} (revealing implementation errors hidden for many years and now fixed), and correctness of an implementation of an encoder and decoder for a popular recent lossless image compression format \cite{BucevKuncak22FormallyQOI}.

\section{A Verified BitStream Implementation}
\label{sec:bitstream}

To introduce the nature of our correctness properties, we present the key mutable data structure and functions for serializing and deserializing basic data types, such as signed and unsigned machine integers.
This data structure, along with the proof of its correctness that we describe in this section, represents around 3700 lines of code. It is invoked by the generated code for serializers of complex data types, whose overall correctness we wish to prove.  

We start with a data structure implementation representing a stream of bits. This data structure is used both as an output medium when serializing data and as input when deserializing.

As the structure needs to offer good runtime performance, it is based on a buffer in the form of an array of bytes (i.e., 8-bit integers). The structure then keeps track of the current index, i.e., the head of the stream, by keeping two variables: \textit{currentByte} and \textit{currentBit}. The \textit{currentByte} value gives the index of the byte in the array where the head is located, and the \textit{currentBit} gives the bit index within this byte. We can then state an invariant that these variables have to satisfy for the structure to be valid:
\begin{lstlisting}
def invariant(currentBit: Int, currentByte: Int, buffLength: Int): Boolean =
  currentBit >= 0 && currentBit < 8 && currentByte >= 0 &&
  ((currentByte < buffLength) || (currentBit == 0 && currentByte == buffLength))
\end{lstlisting}

This invariant states that the \textit{currentByte} points to a byte within the underlying buffer bounds and that the \textit{currentBit} points to a bit within the range of a byte (i.e., between 0 and 7).

Let us now discuss what operations this \textit{BitStream} data structure offers. We divide them into three categories.

First of all, we have functions that work on the index. Some of these functions are moving the head of the stream back and forth, such as \lstinline{increaseBitIndex}, which increases the index by one while taking care of the \textit{currentByte} and \textit{currentBit} arithmetic, or \lstinline{moveBitIndex} which moves the head by a given offset (positive or negative). Let us focus on one of those functions, \lstinline{resetAt}, which has a different use case. It takes another \lstinline{BitStream} as an argument and returns a new instance of \lstinline{BitStream}, with the buffer of the current instance but the head index of the argument \lstinline{BitStream}. This function is used only in specification, and we will discuss how it is used in the section about proofs.

A second type of function offered by the \lstinline{BitStream} class is predicates about the index. These predicates test the available space in the \lstinline{BitStream} after the current head to ensure enough room when encoding or decoding. Several are offered, but they all take size as an argument (in bytes or bits) and return a boolean value indicating whether this space exists between the stream's current head and the buffer's end.

Finally, the \lstinline{BitStream} class offers functions to encode and decode bits to and from the stream. Table \ref{tab:bitstreamFunctions} summarises these functions, showing the encoding and corresponding decoding functions. There are two main classes: the functions encoding and decoding one or more bits and those encoding and decoding one or more bytes. Some functions are implemented by calling other functions in a loop. For example, \lstinline{appendNBits} uses \lstinline{appendBit}.

\begin{table}[ht!]
\centering
\caption{\lstinline{BitStream} Encoding and corresponding Decoding Functions}
\label{tab:bitstreamFunctions}
\ssmall
\begin{tabular}{p{0.6\textwidth} p{0.35\textwidth}}
\toprule
\multicolumn{2}{c}{\textbf{Bit-Level Operations}} \\
\midrule
\textbf{Encoding Functions} & \textbf{Decoding Functions} \\
\midrule
\texttt{appendBit}, \texttt{appendBitOne}, \texttt{appendBitZero} & \texttt{readBit} \\
\texttt{appendNBits}, \texttt{appendNZeroBits}, \texttt{appendNOneBits} & \texttt{readBits} \\
\texttt{appendBitFromByte} & \texttt{readBit} \\
\texttt{appendBitsLSBFirst} & \texttt{readNBitsLSBFirst}\\
\texttt{appendLSBBitsMSBFirst} & \texttt{readNLSBBitsMSBFirst} \\
\texttt{appendBitsMSBFirst} & \texttt{readBits}, \texttt{peekBit} \\
\midrule
\multicolumn{2}{c}{\textbf{Byte-Level Operations}} \\
\midrule
\texttt{appendPartialByte} & \texttt{readPartialByte} \\
\texttt{appendByte} & \texttt{readByte} \\
\texttt{appendByteArray} & \texttt{readByteArray} \\
\bottomrule
\end{tabular}
\end{table}

\subsection{Specification and Verification}

We next present our specifications and formal verification of the functions. We split the specifications into two steps, representing two levels of complexity and guarantees: the absence of crashes (runtime safety) and semantic correctness.

Runtime safety corresponds to the program not crashing. It ensures that all array accesses are within bounds, that integer operations do not overflow, and that no division by zero can occur. To be able to prove the in-bound accesses of arrays, we prove for each function which encodes or decodes data to or from the stream that it is moving the current head by the required number of bits, and we add as a precondition that the number of available bits is greater or equal to the size that will be written or consumed from the stream. For example, for the function \lstinline{appendBitsMSBFirst}, we prove that the head index after the call is equal to the head index before plus \lstinline{nBits}, which is the number of bits written by the function. To prove the absence of overflows and division by zero, we add preconditions on arguments and some runtime sanitization checks when the property cannot be ensured statically. For example, these properties ensure the number of bits to encode or decode received as arguments is greater than zero and not too large or that indices are greater or equal to zero and smaller than the biggest integer value.

Once the runtime safety is guaranteed, we can prove some semantic properties. In this case study, the property of interest is the \textit{invertibility}. In the context of the \lstinline{BitStream} class, it can be intuitively summarized by saying that after encoding some value, reading from the stream returns the same value. This second part of the specification is more challenging to prove. More specifically, the postcondition of the encoding function states that the return value of a call to their corresponding decode function (see Table \ref{tab:bitstreamFunctions}) is equal to what has been written. For some functions, what is written is not exactly the input value. For example, \lstinline {appendBitFromByte} writes only one bit of the received byte, so the postcondition states only that the written bit is the expected one.

We explore in more detail the proof of the function \lstinline{appendBitsMSBFirst}, which has the most interesting proof and was the most challenging to verify. First, here is the function's implementation and postcondition part about invertibility (note that the precondition, proof lines, and parts of the postcondition are omitted):

\begin{lstlisting}{scala}
def appendBitsMSBFirst(srcBuffer: Array[UByte], nBits: Long, from: Long = 0): Unit = {
    appendBitsMSBFirstLoop(srcBuffer, from, from + nBits)
}.ensuring: // ...
  val (r1, r2) = reader(w1, w2) // returns a new bitstream with the buffer
                                // of w1, but the head of w2
  validateOffsetBitsContentIrrelevancyLemma(w1, w2.buf, nBits)
  val vGot = r1.readBits(nBits)
  byteArrayBitContentSame(srcBuffer, vGot, from, 0, nBits) // Compare the bit
  // content of the two arrays between from and from + nBits, respectively 0 and nBits.
  
def appendBitsMSBFirstLoop(srcBuffer: Array[UByte], i: Long, to: Long): Unit = {
   if i < to then
      appendBitFromByte(srcBuffer((i / 8).toInt).toRaw, (i % 8).toInt)
      appendBitsMSBFirstLoop(srcBuffer, i + 1, to)
}.ensuring:
   val (r1, r2) = reader(old(this), this)
   validateOffsetBitsContentIrrelevancyLemma(old(this), this.buf, to - i)
   val listBits = bitStreamReadBitsIntoList(r1, to - i)
   val srcList = byteArrayBitContentToList(srcBuffer, i, to - i)
   listBits == srcList
\end{lstlisting}

The function is effectively a loop calling \lstinline{appendBitFromByte} for each index $i$ between \lstinline{from} and \lstinline{from + nBits}, writing the $i^{th}$ bit to the stream. To aid verification, we implemented the loop as a tail-recursive function (Scala compiler transforms such recursion back to a while loop, so no efficiency is lost).

To give a sense of the size of the proof compared to the size of the code, this function comprises less than 10 lines of code. However, this number increases to 150 when accounting for the proof annotations without even considering the implementations of the different lemmas. If we include those, this number increases further to almost 300 lines of code. This function was one of the most challenging to verify and consequently has one of the worst ratio lines of code versus lines of proof annotations.

The corresponding decode function is \lstinline{readBits}, which reads a given number of bits from the stream and returns them in an array of bytes. This function is also implemented as a tail-recursive function calling \lstinline{readBit}.

\begin{figure}[!h]
    \centering
    \includegraphics[width=0.88\textwidth]{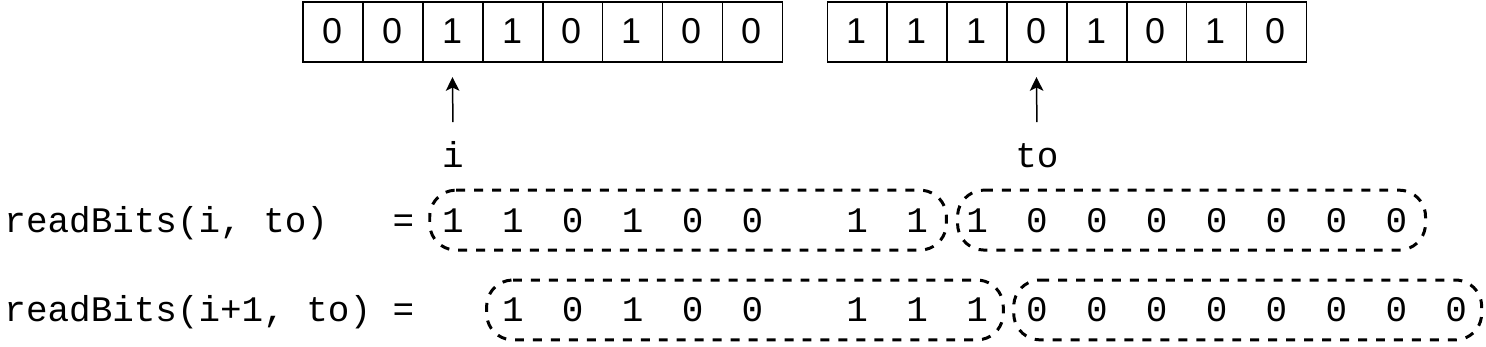}
    \caption{Applying \lstinline{readBits} at \lstinline{i} and \lstinline{i+1} yields different arrays.}
    \label{fig:readbits}
\end{figure}

Proof of invertibility is by induction. The induction hypothesis at iteration $i$ states that the recursive call will correctly encode the bits from $i + 1$ to $to$, meaning that the postcondition is correct. It is then enough to prove that encoding the bit $i$ is correct, i.e., the \lstinline{readBits} function $i^{th}$ iteration would correctly read the bit $i$. However, the data are represented by arrays of bytes, not lists of bits, which makes the proof more complicated. Indeed, decoding the bits from $i + 1$ to $to$ gives an entirely different array of bytes from decoding $i$ to $to$, as all bits are shifted. Figure~\ref{fig:readbits} illustrates this mismatch.

\begin{figure}[!h]
    \centering
    \includegraphics[width=\textwidth]{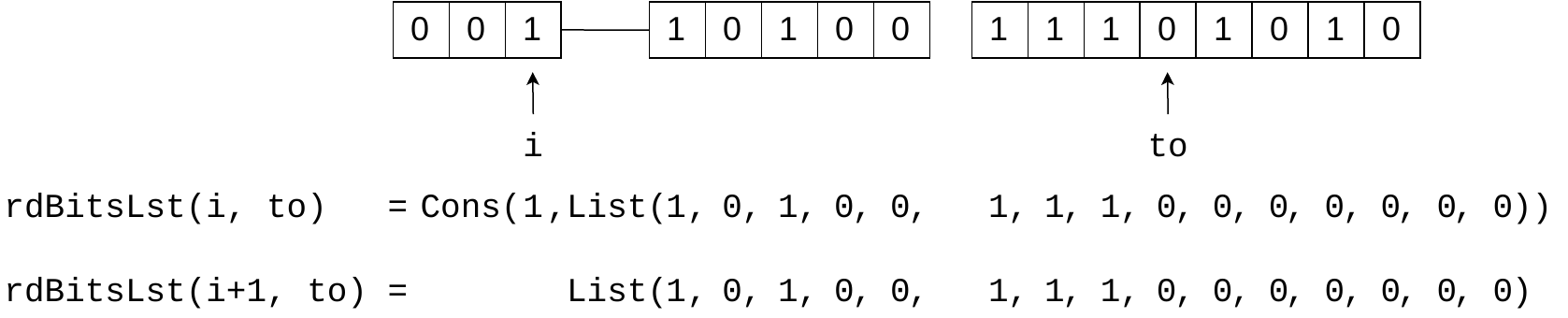}
    \caption{The result of \lstinline{bitStreamReadBitsIntoList} (abbreviated \lstinline{rdBitsLst}) at \lstinline{i+1} is a suffix of the resulting list at \lstinline{i}.}
    \label{fig:readbitslst}
\end{figure}

Applying the induction hypothesis in this context is not automatically possible. We, therefore, wrote a proof using a detour through the list of bits. Concretely, we implemented two functions: \lstinline{bitStreamReadBitsIntoList} which reads \lstinline{nBits} from a given \lstinline{BitStream} instance and returns a list of boolean values, and \lstinline{byteArrayBitContentToList} which transforms a list of booleans into an array of bytes. Proceeding with the previous example, Figure~\ref{fig:readbitslst} shows the result of decoding using the function returning a list of booleans.

We can see that the result as a list has a form that makes applying the induction hypothesis possible. Indeed, the list at iteration $i$ equals the list of the result of the recursive call, with the $i^{th}$ bit prepended. At that stage, the specification of \verb|appendBitsMSBFirstLoop| ensures that the content of the bitstream after the call, read as a list by \verb|bitStreamReadBitsIntoList|, is equal to the content of the \verb|srcBuffer| also read as a list. We then write a lemma to prove that two arrays of bytes for which \verb|bitStreamReadBitsIntoList| returns the same list have the same content when compared with \verb|byteArrayBitContentSame|. This lemma can be applied when calling \verb|appendBitsMSBFirstLoop| in \verb|appendBitsMSBFirst| to finalize the proof.
\section{Codecs}
\label{sec:codecs}

ASN1SCC relies on static sets of functions that encode and decode some basic datatypes on and from a \lstinline{BitStream} instance based on some format. These sets of functions are called \textit{codecs}, and there are three different codecs in this project, corresponding to three formats: UPER, PER, and ACN. We focus on the ACN codec as this is the one used by the generated code we verify in Section~\ref{sec:proofgen}.

Codecs are implemented as decorators of the \lstinline{BitStream} class, i.e., their interfaces are a superset of the \lstinline{BitStream} interface. This means that each instance of a codec contains a \lstinline{BitStream} instance and forwards function calls to it for functions of the \lstinline{BitStream} interface. The common functionality between the three abovementioned codecs is extracted in a \lstinline{Codec} class. Therefore, the \lstinline{Codec_ACN} class contains an instance of \lstinline{Codec} itself, which, in turn, contains an instance of \lstinline{BitStream}. The following is an example of a pair of functions implemented in \lstinline{Codec}:
\begin{lstlisting}
def encodeConstrainedPosWholeNumber(v: ULong, min: ULong, max: ULong): Unit
def decodeConstrainedPosWholeNumber(min: ULong, max: ULong): ULong
\end{lstlisting}
These functions encode (respectively decode) an unsigned 64-bit integer in the interval $[min, max]$.

The complete list of functions implemented in \lstinline{Codec_ACN} used in the code generated by the ASN1SCC compiler is shown in Table \ref{tab:acnCodecsFunctions}. The common class \lstinline{Codec} and the specific class \lstinline{Codec_ACN}, along with the proof annotations, comprise around 4000 lines of code.

We performed similar verification work as for the \lstinline{BitStream} class  described in Section \ref{sec:bitstream}. We proved runtime safety for all of the functions listed in Table \ref{tab:acnCodecsFunctions} and proved invertibility for most of them. We did not prove the invertibility of the functions working with \lstinline{Real} numbers (following the IEEE754 norm) because Stainless does not support floating point numbers (unlike a related Daisy tool \cite{DBLP:conf/sas/IsychevD23}). We also did not yet prove the invertibility of String encoding and decoding functions.

The approach used to write the proof is the same as for the \lstinline{BitStream} class (explained in Section \ref{sec:bitstream}) and relies a lot on the properties proved on \lstinline{BitStream}'s operations.

While writing the proof of invertibility, we performed an interesting refactoring to one auxiliary function: \lstinline{uint2int}. This function converts an unsigned integer of a given number of bytes to a 32-bit signed integer. The unsigned integer is a 64-bit integer in which only the given number of bytes are considered. This function was implemented using a \lstinline{while} loop to iterate over the number of bytes. Stainless supports verification of \lstinline{while} loops, but verifying functions that use them requires writing invariants that can be complex and, therefore, hard to come up with. A basic invariant was insufficient to prove the needed properties in this case. Instead of trying to find a sufficient invariant, given that the number of iterations the loop could perform was bound by 0 and 7, we decided to unroll the loop by hand in the source code. While preserving the code's conciseness, this manual unrolling allows automatic verification of properties that would have otherwise required a non-trivial invariant. 
This illustrates that refactoring can simplify verification while preserving the semantics.

\begin{table}[ht!]
\centering
\caption{Encoding and Corresponding Decoding Functions}
\label{tab:acnCodecsFunctions}
\ssmall
\begin{tabular}{p{0.49\textwidth} p{0.49\textwidth}}
\toprule
\multicolumn{2}{c}{\textbf{ACN Codec}} \\
\midrule
\textbf{Encoding Function} & \textbf{Decoding Function} \\ \hline
\texttt{enc\_Int\_PositiveInteger\_ConstSize\_8} & \texttt{dec\_Int\_PositiveInteger\_ConstSize\_8} \\ \hline
\texttt{enc\_Int\_PositiveInteger\_ConstSize\_big\_endian\_\{16,32,64\}} & \texttt{dec\_Int\_PositiveInteger\_ConstSize\_big\_endian\_\{16,32,64\}} \\ \hline
\texttt{enc\_Int\_PositiveInteger\_ConstSize} & \texttt{dec\_Int\_PositiveInteger\_ConstSize} \\ \hline
\texttt{enc\_Int\_TwosComplement\_ConstSize\_8} & \texttt{dec\_Int\_TwosComplement\_ConstSize\_8} \\ \hline
\texttt{enc\_Int\_TwosComplement\_ConstSize\_big\_endian\_\{16,32,64\}} & \texttt{dec\_Int\_TwosComplement\_ConstSize\_big\_endian\_\{16,32,64\}} \\ \hline
\texttt{enc\_Real\_IEEE754\_\{32,64\}\_big\_endian} & \texttt{dec\_Real\_IEEE754\_\{16,32,64\}\_big\_endian} \\ \hline
\texttt{enc\_Real\_IEEE754\_\{16,32,64\}\_little\_endian} & \texttt{dec\_Real\_IEEE754\_\{16,32,64\}\_little\_endian} \\ \hline
\texttt{enc\_String\_Ascii\_Null\_Terminated\_multVec} & \texttt{dec\_String\_Ascii\_Null\_Terminated\_multVec} \\ \hline
\texttt{enc\_String\_CharIndex\_private} & \texttt{dec\_String\_CharIndex\_private} \\ \hline
\texttt{enc\_IA5String\_CharIndex\_External\_Field\_DeterminantVec} & \texttt{dec\_IA5String\_CharIndex\_External\_Field\_DeterminantVec} \\ \hline
\texttt{enc\_IA5String\_CharIndex\_Internal\_Field\_DeterminantVec} & \texttt{dec\_IA5String\_CharIndex\_Internal\_Field\_DeterminantVec} \\ \hline
\end{tabular}
\end{table}

\section{Tailoring ASN1SCC to Verification}
We next present our design of a generator that produces verifiable code along with inductive specifications, while preserving the spirit of existing C and Ada generators. 
Our approach illustrates that verification is best done in a situation when it is possible to adjust the code, balancing efficiency and verifiability.
We first discuss the decoders. Whereas C and Ada decode the result in-place,
we redesigned the Scala backend to construct fresh values. We review below the replacement of mutable arrays
by immutable sequences, the alternatives we considered, and the use of recursion instead of imperative loops.
We then consider the generation of alignment-aware size methods for \lstinline{SEQUENCE}, \lstinline{SEQUENCE OF} and \lstinline{CHOICE}.
We finally examine the case of ACN-specialized type assignments, the result in the generated code,
their negative impact on verification, and the countermeasures.

\subsection{Alternatives to In-Place Mutation}
ASN1SCC generated code for existing targets (C and Ada) leverages in-place mutation for decoding.
The caller of any decoding function is expected to provide a structure.
Arrays within structures (e.g. strings or \lstinline{SEQUENCE OF}) are expected to be allocated with the maximum size.
As the decoding progresses, the structure is updated with the decoded values.
Of particular interest is decoding a \lstinline{CHOICE}.
The C template leverages \lstinline{union}. Since these are unchecked, they can be written without additional constraints.
Assuming we are decoding a \lstinline{SEQUENCE} named \lstinline{S} in an instance \lstinline{s} with a field \lstinline{choice} whose variant is decided by \lstinline{kind}, the generated C would be:

\begin{lstlisting}[language=c]
union { C1 c1; /* ... */ Cn cn;} C_union; struct { C_selection kind; C_union u; } C;
struct { /* ... */ C choice } S;
switch (kind) {
  case C1_tag: s->choice.kind = C1_tag; decodeC1(&s->choice.u.c1, bitStream);
  // ...
  case Cn_tag: s->choice.kind = Cn_tag; decodeCn(&s->choice.u.cn, bitStream);}
\end{lstlisting}
In the above snippet, the field \lstinline{kind} stores the tag of the \lstinline{CHOICE} variant, and the field \lstinline{u} is a union of all \lstinline{CHOICE} variants.

The Ada template uses variant records. It deviates from the C backend because it must first default-initialize the variant.
The rest, however, remains similar. In particular, the variant is selected and decoded in place.

This template where the choice variant is selected and mutated is not possible in Scala, as it results in a typing error:

\begin{lstlisting}
enum C { case C1(...); /* ... */ case Cn(...) }
kind match
  case C1_tag => s.choice=C1_Init(); decodeC1(s.choice, bitStream)//expect C1, got C
  case Cn_tag => s.choice=Cn_Init(); decodeCn(s.choice, bitStream)//expect Cn, got C
\end{lstlisting}

A solution is to declare and initialize a local variable, assign it to \lstinline{s.choice} and pass it to the decoding function.
Unfortunately, Stainless does not allow this kind of mutation as it is not part of the aliasing fragment it supports.
This aspect motivated the change of the Scala decoding functions to return decoded values instead of relying on mutation. For instance, the above example becomes:

\begin{lstlisting}
val choice = kind match
  case C1_tag => decodeC1(bitStream); case Cn_tag => decodeCn(bitStream)
S(..., choice, ...) // Return the decoded S with the decoded choice and other fields
\end{lstlisting}

When decoded, individual elements of \lstinline{SEQUENCE OF} are also returned as values.
There are, however, two possible ways to decode the overall \lstinline{SEQUENCE OF}.
One way is to create a local array of the appropriate size, update it in place, and return it.
Another one is to append the decoded elements to a collection. This possibility can be split depending on whether we choose the collection to be immutable or not.
We detail the approach we retain in the following section.

\subsection{Data Structures Representing \lstinline{SEQUENCE OF}}
ASN1SCC employs mutable arrays to store the elements of a \lstinline{SEQUENCE OF}.
For decoding, the array is pre-allocated to the maximum size stated in the ASN.1 specification.
Individual elements are decoded in place by dereferencing the element for both the C and Ada backends.

Due to our choice of decoding elements (relying on constructing values instead of mutating them in place, as discussed earlier),
the template for decoding individual \lstinline{SEQUENCE} OF elements also needs to be adapted.
One solution would be to create a local array and update it with the decoded elements:

\begin{lstlisting}
val arr = Array.fill(len)(elemInit()) // Default-initialized elements
for (i <- 0 until len) { val elem = decode(bitStream); arr(i) = elem }
S(..., arr, ...) // Return the structure wrapping the array with other decoded fields
\end{lstlisting}

Stainless will accept this snippet as long as the type of the element does not contain any other \lstinline{Array}.
If, however, the type transitively contains an \lstinline{Array}, we will stumble upon the same issue of unsupported aliasing, as mentioned earlier.
It is possible to create an alias analysis escape hatch.
We need to uphold the unchecked assumption that the decoded elements are not mutated for the rest of the program.
While suboptimal, this could be considered an acceptable solution because the code we generate will maintain this invariant.
Such a solution still fails, however, when it comes to actual verification conditions. Namely, Stainless encodes the JVM \lstinline{Array} leveraging the SMT theory of generalized arrays with map combinators \cite{DBLP:conf/fmcad/MouraB09}, a \lstinline{SEQUENCE OF} is therefore represented in part as an SMT array.
The encoding becomes more complex in the presence of nested \lstinline{SEQUENCE OF} and uses a non-standard array combinator.
While Z3 supports this combinator, it sometimes times out for intermediate queries. On the other hand, cvc5 does not support it.

Our solution is to wrap a Scala \lstinline{Vector} in a class that exposes commonly used operations. ASN1SCC relies on random access on some procedures, which is an acceptable $O(\log{n})$ for \lstinline{Vector}. 
We specify vector operations using a \lstinline{List}.
This specification has the advantage of relying on the SMT theory of ADT.
The Stainless library provides many properties and lemmas regarding operations on \lstinline{List}. Our solution thus achieves a usable combination of efficiency and verifiability.

We also lift the recursive function for encoding and decoding a \lstinline{SEQUENCE OF}
to the top-level. This enables the encoding function to refer to its decoding counterpart in the postcondition, to specify the invertibility property.

\subsection{Size Computation}
\label{subsec:size-comp}
ASN1SCC computes a lower and an upper bound of their size in bits for each type assignment.
These statically known bounds are sufficient to prove runtime safety.
For invertibility, however, the exact position of the bitstream must be known.
This, in turn, requires computing the exact size of structures, some of which may be dynamic (e.g. variable-sized strings, \lstinline{SEQUENCE OF} or \lstinline{CHOICE}).

We generate size methods for all class definitions.
ACN allows fields to have an alignment restriction, impacting the size of the overall structure.
Therefore, the size of such a structure depends on the bitstream position.
The generated size methods take an offset corresponding to the bitstream position as an argument to account for this.
The size of structures with no alignment restriction in any of their component is invariant in the bitstream position.
Structures having a component with a byte alignment restriction and no word (16 bits) or double word alignment (32 bits) have a size invariant in the offset modulo 8, 16 and 32. A similar observation can be made for word (invariant modulo 16 and 32) and double word alignment (invariant modulo 32).
We generate lemmas stating these properties and apply them whenever the bitstream must align its cursor.
The size methods are implemented by recursively computing the size of each element and threading the offset with the accumulated size.

\subsection{The Case of ACN-Specialization}
\label{subsec:acn-spec}
For type assignments whose binary format is described with ACN, ASN1SCC will specialize the encoding and decoding functions.
It does so by inlining calls to these type assignments' encoder and decoder functions and by inserting the necessary logic for the ACN-specific part.
This specialization by inlining does not hinder verification much for small structures and simple properties.
However, for invertibility, this becomes an issue because the verification conditions (VCs) are already complex and large.
We solve this problem by ``restoring'' some modularity and hoisting (or outlining) the inlined code into top-level functions.
We also parameterize the functions by their ACN dependencies and the ACN fields they return.
\section{Proof Generation for \lstinline{SEQUENCE} Invertibility}
\label{sec:proofgen}
When invoked with the \lstinline{-invertibility} flag, our code generator produces additional postconditions, assertions and lemmas to prove \lstinline{SEQUENCE} invertibility.
The extra postconditions essentially state that decoding the encoded message yields the original result.
These are also generated for \lstinline{CHOICE} and \lstinline{SEQUENCE OF}, even though currently, no proofs are produced for the invertibility of these recursive cases.

For a structure \lstinline{s} (whether a \lstinline{SEQUENCE}, \lstinline{CHOICE} or \lstinline{SEQUENCE OF}), the postcondition is as follows:
\begin{lstlisting}
@opaque def S_Encode(s: S, codec: ACN) = { /* ...implementation... */ }.ensuring:
  case Left(_) => true // 1.
  case Right(_) => // 2.
    old(codec).buf.length == codec.buf.length && // 3.
    codec.bitIndex == old(codec).bitIndex + s.size(old(codec).bitIndex) && // 4.
    old(codec).isPrefixOf(codec) && locally: // 5.
      val r1 = codec.resetAt(old(codec)) // 6.
      val (r2Got, decRes) = S_Decode_pure(r1) // 7.
      decRes match
        case Left(_) => false // 8.
        case Right(resGot) => r2Got == codec && resGot == s // 9.
\end{lstlisting}
At line 1, no properties are stated if the encoding fails.
If it succeeds (line 2), line 3 states that the \lstinline{codec} before entering this function (denoted by \lstinline{old(codec)}) has its buffer length unchanged.
Line 4 states that the cursor of the codec is precisely advanced by the size of \lstinline{s}.
As discussed previously, a structure may have some alignment restriction in one or multiple of its fields.
Its overall size, therefore, depends on the starting position of the bitstream, which we pass to \lstinline{s.size}.
These stated properties are always generated, even if \lstinline{-invertibility} is not passed.
Line 5 essentially indicates that the function did only append in the buffer (hence, the old codec is a prefix of the new one).
The condition wrapped in a local block starting at line 5 is the inversion property.
In line 6, we rewind the bitstream back to the original position.
We then decode what we encoded using a pure version of the corresponding decoding function at line 7.
It conceptually makes a copy of the given bitstream and returns the mutated version in \lstinline{c2Res} and the decoded result in \lstinline{decRes}.
Line 8 states that the decoding cannot fail, and line 9 specifies that the decoded value is the same as we started with.

The same template applies to ACN-specialized hoisted functions, discussed earlier.
The difference lies in the additional parameters, and the returned decoded ACN fields (if any).
The invertibility accounts for these differences.
In particular, it ensures that the decoded ACN fields are equal to what was encoded.

We furthermore annotate the encoding function with \lstinline{@opaque}, effectively hiding the body to the solver, as the postcondition exactly states the behavior of the function.
This improves the verification performance since the solver does not need to unfold the body (which, in turn, would lead to unfolding other encoding calls).
To improve performance further, we also annotate the decoding functions as opaque.
The postcondition of the decoding functions is incomplete (they only state properties 1 through 4); therefore, we need to explicitly unfold them when proving their invertibility (inside the encoding function) or the lemmas related to them.
Failing to do so would lead to ``false'' counterexamples.

Proving \lstinline{SEQUENCE} invertibility relies on the composition of the invertibility of each field but is insufficient on its own.
The postcondition for a field \lstinline[mathescape=true]{f$_i$} only states that decoding from the bitstream at $i+1$ (rewound at $i$) yields the same result, while we need it to hold for the bitstream at $n+1$ rewound at $i$.
We intuitively need a ``prefix lemma'' stating that if one bitstream is a prefix of another one up to the size of the structure to decode, then the results are equivalent.
The size of the structure to decode is unknown in the context of this lemma, we need to correctly ``guess'' it (which we detail next).
When applying the lemma, the size simply corresponds to the size of the field we encoded, which we know.

The template of a prefix lemma follows. It applies to \lstinline{SEQUENCE}, \lstinline{CHOICE}, and \lstinline{SEQUENCE OF}. Only \lstinline{SEQUENCE OF} does not have a proof.
ACN-specialized decoding functions may have additional parameters and are accounted for.

\begin{lstlisting}
def T_prefixLemma(c1: ACN, c2: ACN, sz: Long): Unit =
  require(c1.buf.length == c2.buf.length && c1.validate_offset_bits(T_MAX_SIZE) &&
    0L <= sz && sz <= T_MAX_SIZE) // 1.
  require(arrayBitRangesEq(c1.buf, c2.buf, 0L, c1.bitIndex + sz)) // 2.
  val c2Rst = c2.resetAt(c1) // 3.
  val (c1Res, decRes1) = T_Decode_pure(c1) // 4.
  val (c2Res, decRes2) = T_Decode_pure(c2Rst) // 5.
  val v1Size = decRes1 match // 6.
    case Right(v1) => v1.size(c1.bitIndex); case Left(_) => 0L
  { /* proof */ }.ensuring:
    decRes1 match
      case Right(v1) if v1Size == sz => // 7.
        decRes2 match
          case Right(v2) => c1Res.bitIndex == c2Res.bitIndex && v1 == v2 // 8.
          case Left(_) => false // 9.
      case Left(_) => true // 10.
\end{lstlisting}

The lemma assumes the following. The two codecs must have the same buffer length.
\lstinline{c1} needs sufficient space to decode any \lstinline{T}, with \lstinline{T_MAX_SIZE} being the statically known maximum size for any instance of \lstinline{T}.
The size parameter must be positive and no greater than \lstinline{T_MAX_SIZE}.
\lstinline{c1} and \lstinline{c2} must be equal in content up the position of \lstinline{c1} plus the size to be ``guessed''.
Then, the lemma states the following.
We rewind \lstinline{c2} at the position of \lstinline{c1} at line 3. Line 4 and 5 decode a \lstinline{T} with \lstinline{c1}, and \lstinline{c2} rewound at \lstinline{c1}.
The resulting codecs are stored in \lstinline{c1Res} and \lstinline{c2Res} respectively.
The decoding results are bound to \lstinline{decRes1} and \lstinline{decRes2}.
We compute the size of the decoded message from \lstinline{c1} at line 6. In case of failure, it is arbitrarily set to 0.
The lemma result is stated in the ensuring clause.
If decoding from \lstinline{c1} fails or if the guessed size is different from the resulting size, there are no claims (line 10).
Otherwise (line 7), the lemma states that decoding from \lstinline{c2} cannot fail (line 9).
Furthermore, the decoded value is the same as the one from \lstinline{c1}, and the codecs end up in the same position (line 8).

Proving this property is a matter of applying the corresponding prefix lemma for each field along with other lemmas.
Without further proof engineering, performance unfortunately greatly suffers even with the outlining of ACN-specialized function.
The main culprit is in the need to unfold both the body of \lstinline{T_Decode(c1)} and \lstinline{T_Decode(c2Rst)} at the beginning of the proof.
This is necessary to prove that decoding \lstinline[mathescape=true]{f$_i$} cannot fail thanks to the end-to-end decoding success.
Additionally, many intermediate assertions cause the VCs for subsequent fields to grow.

As a first step to make verification feasible, we wrap the proof for each field in a local opaque function and state the desired property in the postcondition.
We apply these sublemmas sequentially, then unfold the body of \lstinline{T_Decode} applied with \lstinline{c1} and \lstinline{c2Rst} to ``glue'' everything together.
The proof becomes:

\begin{lstlisting}[mathescape=true]
decRes1 match
  case Right(v1) if v1Size == sz =>
    @opaque def proof_f$_i$(c1$_i$: ACN, c2$_i$: ACN): Unit = {
      require(c1$_i$.buf == c1.buf && c2$_i$.buf == c2.buf) // 1.
      val offset = size$_1$ + ... + size$_{i-1}$ // 2.
      require(c1$_i$.bitIndex == c1.bitIndex+offset && c1$_i$.bitIndex == c2$_i$.bitIndex)// 3.
      arrayBitRangesEqSlicedLemma(c1$_i$.buf, c2$_i$.buf,
        0L, c1$_i$.bitIndex + v1Size - offset, 0L, c1$_i$.bitIndex + size$_i$) // 4.
      // ... assertions to prove requirements ...
      T$_i$_prefixLemma(c1$_i$, c2$_i$.withMovedBitIndex(size$_i$), size$_i$) // 5.
      // ... other assertions
    }.ensuring:
      val (c1$_i$+1, dec1) = T$_i$_Decode_pure(c1$_i$) // 6.
      val (c2$_i$+1, dec2) = T$_i$_Decode_pure(c2$_i$) // 7.
      val f$_i$_1 = dec1 match case Right(res) => res; case Left(_) => ??? // 8.
      val f$_i$_2 = dec2 match case Right(res) => res; case Left(_) => ??? // 9.
      f$_i$_1.size(c1$_i$.bitIndex) == size$_i$ && f$_i$_1 == f$_i$_2 && c1$_{i+1}$.buf == c1.buf &&
      c2$_{i+1}$.buf == c2.buf && c1$_{i+1}$.bitIndex == c1.bitIndex + offset &&
      c1$_{i+1}$.bitIndex == c2$_{i+1}$.bitIndex // 10.
    unfold(T_Decode(snapshot(c1))) // 11.
    proof_f$_i$(c1$_i$, c2$_i$)
    val(c1$_{i+1}$, dec1$_{i+1}$)=T$_i$_Decode_pure(c1$_i$);val(c2$_{i+1}$, dec2$_{i+1}$)=T$_i$_Decode_pure(c2$_i$)
    unfold(T_Decode(snapshot(c2Rst))) // 12.
    decRes2 match
      case Right(v2) => check(c1Res.bitIndex == c2Res.bitIndex && v1 == v2)
      case LeftMut(_) => check(false)
  case _ => () // vacuous
\end{lstlisting}

For each field \lstinline[mathescape=true]{f$_i$}, we generate a corresponding \lstinline[mathescape=true]{proof_f$_i$}.
It takes two codecs \lstinline[mathescape=true]{c1$_i$} and \lstinline[mathescape=true]{c2$_i$} corresponding to the outer \lstinline{c1} and \lstinline{c2} rewound at the offset of \lstinline[mathescape=true]{f$_i$}, conditions stated in 1 and 3.
The offset computed at 2 refers to the sizes of each previous field from the decoded \lstinline{v1} and is computed outside of this function.
For $i = 1$, \lstinline[mathescape=true]{proof_f$_1$} is parameterless and directly uses \lstinline{c1} and \lstinline{c2Rst}.
Application of the lemma at 4 allows to deduce that \lstinline[mathescape=true]{c1$_i$} and \lstinline[mathescape=true]{c2$_i$} are equal in [0, \lstinline[mathescape=true]{c1$_i$}.bitIndex + \lstinline[mathescape=true]{size$_i$})
where \lstinline[mathescape=true]{size$_i$} is the size of \lstinline[mathescape=true]{f$_i$}. This property is needed to be able to apply the prefix lemma at 5.
\lstinline[mathescape=true]{proof_f$_i$} states that decoding from \lstinline[mathescape=true]{c1$_i$} and \lstinline[mathescape=true]{c2$_i$} (lines 6 and 7) cannot fail (lines 8 and 9) and that the values are equal, alongside other properties (line 10).
Intuitively, proving that decoding from \lstinline[mathescape=true]{c1$_i$} succeeds requires unfolding the body of \lstinline{T_Decode(c1)} in order to unveil that this step is part of \lstinline{T_Decode}.
It is, however, insufficient since the \lstinline[mathescape=true]{c1$_i$} of \lstinline[mathescape=true]{proof_f$_i$} has no correspondence with the state of the codec in \lstinline{T_Decode} at $i$ (having the same position and buffer is insufficient).
We discuss one solution afterward.
On the other hand, \lstinline[mathescape=true]{T$_i$_prefixLemma} guarantees that decoding from \lstinline[mathescape=true]{c2$_i$} cannot fail provided decoding from \lstinline[mathescape=true]{c1$_i$} succeeds.

The actual proof of \lstinline{T_prefixLemma} starts at 11. We unfold \lstinline{T_Decode} applied with \lstinline{c1}. Note that unfolding \lstinline{T_Decode_pure} is futile because it calls \lstinline{T_Decode}, which will not be unfolded.
That said, \lstinline{T_Decode} mutates its parameter. We therefore need to make a copy of \lstinline{c1} (done with \lstinline{snapshot}).
For each field \lstinline[mathescape=true]{f$_i$}, we then apply the corresponding proof and call the decoding function. We thread the resulting codecs to the following field.
For ACN-specialized outlined functions, functions returning ACN fields have their value extracted and forwarded where needed.
At line 12, we unfold \lstinline{T_Decode} applied with \lstinline{c2Rst} to prove that decoding from \lstinline{c2} (rewound at \lstinline{c1}) cannot fail and yield the same values.

Going back to proving the infallibility in \lstinline{proof_f$_i$}, a solution consists in first creating an ``origin function'' which threads the various codecs:
\begin{lstlisting}
@opaque def f$_i$_codec_origin(c1$_i$: ACN): Boolean =
  val (c1$_2$_got, dec$_2$_got) = T$_1$_Decode_pure(c1)
  dec$_2$_got match
    case Right(_) => // ... decode T$_2$ and so on
      val (c1$_i$_got, dec$_i$_got) = T$_i$_Decode_pure(c$_{i-1}$)
      dec$_i$_got match case Right(_) => c1$_i$_got == c1$_i$; case Left(_) => false
    case Left(_) => false
\end{lstlisting}

Then, we add \lstinline[mathescape=true]{f$_i$_codec_origin(c1$_i$)} as a precondition to \lstinline[mathescape=true]{proof_f$_i$} and unfold it along with \lstinline{T_Decode} to prove infallibility.
\section{Experience with Case Study}
\label{sec:results}

\subsection{Verified Properties and Statistics}
\label{subsec:ver-props}
\subsubsection{Codec classes}
\label{subsec:ver-codecs}
We first present the verification of the \lstinline{BitStream} and codecs classes in Table~\ref{tab:vcs}, column \textit{Library}.
Runtime safety is proven for all functions. Invertibility is proved for all functions used by the PUS-C services, minus the floating point and string-related functions.
We conducted the experiment with a timeout of 6 minutes, with Z3 v4.12.2, cvc5 1.1.2 and CVC4 1.8.

We report for each VC category the number of \textbf{V}erified, \textbf{U}ndecided, and \textbf{I}nvalid VCs.
\emph{Measures} VCs ensure that the measure annotations for recursive functions and \lstinline{while} loops are positive and strictly decreasing.
\emph{Class invariants} check that instances of classes with invariant uphold
the stated properties (whether on construction or on mutation). It mostly relates to updates of codec instances.
\emph{Pos(itive) array size} refers to VCs checking array allocation size to be positive.
\emph{Miscellaneous} contains all VCs introduced by some Stainless phases.

\begin{table}[!htb]
\centering
\caption{Statistics of Verification Conditions.}
\begin{tabular}{l r r r|r r r|r r r}
                  & \multicolumn{3}{c|}{Library} & \multicolumn{3}{c|}{PUS-C services} & \multicolumn{3}{c}{\lstinline{TC-Packet}} \\
VCs               & \# V & \# U & \# I & \# V & \# U & \# I & \# V & \# U & \# I \\
\hline
Preconditions     &   4,252 &       0 &       0 & 152,201 & 1       &       2 &     529 &       0 &        0 \\
Overflows/casts   &     936 &       0 &       0 &  82,037 & 0       &       0 &     230 &       0 &        0 \\
Assertions        &     544 &       0 &       0 &  23,284 & 0       &       0 &     167 &       0 &        0 \\
Postcondition     &     443 &       0 &       0 &  22,365 & 1       &       0 &      30 &       0 &        0 \\
Arithmetic ops    &     183 &       0 &       0 &   3,711 & 0       &       0 &       0 &       0 &        0 \\
Array access      &     181 &       0 &       0 &       0 & 0       &       0 &       0 &       0 &        0 \\
Measures          &     132 &       0 &       0 &   2,796 & 0       &       0 &       0 &       0 &        0 \\
Class invariant   &      54 &       0 &       0 &   1,722 & 0       &       0 &       0 &       0 &        0 \\
Match exh.        &      39 &       0 &       0 &  38,283 & 0       &       0 &     101 &       0 &        0 \\
Pos. array size   &       5 &       0 &       0 &       0 & 0       &       0 &       0 &       0 &        0 \\
Miscellaneous     &       2 &       0 &       0 &     918 & 0       &       0 &       0 &       0 &        0 \\
\hline
\hline
\textbf{Total} & \textbf{6,771} & \textbf{0} & \textbf{0} & \textbf{327,317} & \textbf{2} & \textbf{2} & \textbf{1,057} & \textbf{0} & \textbf{0}
\end{tabular}
\label{tab:vcs}
\end{table}

\subsubsection{Generated Scala code}
\label{subsec:ver-genai-i-meant-genscala}
We present two sets of results reported in Table~\ref{tab:vcs}. The first one, column \textit{PUS-C services}, is a run over all but one PUS-C services without the \lstinline{-invertibility} option.
The verified properties include runtime safety, exact structure size computation, and precise bitstream position.
We excluded PUS-C service number 4 because it uses floating point arithmetic, which Stainless does not support.
Additionally, all services share the ASN.1/ACN definitions within \lstinline{ccsds} and \lstinline{common}.
Consequently, these VCs are counted multiple times.
We did not, however, deduplicate them because the different services had different numbers, which was caused by some light variation in the generated code.
When generated and verified in isolation, they have 11,268 VCs, all valid.

The second set of results is the verification of \lstinline{TC-Packet} with the \lstinline{-invertibility} flag and is reported in column \textit{\lstinline{TC-Packet}}.
This \lstinline{SEQUENCE} is the largest in all services.
Note that the individual components are not reported since they contain structures for which no proofs are generated at the moment of writing (such as \lstinline{CHOICE} or \lstinline{NullType} with special encoding).

It should be noted that no in-bound array access VCs are present in both columns due to the usage of the \lstinline{Vector} class: in-bound access is instead verified as a precondition of the indexing method \lstinline{apply}.
Furthermore, the underlying buffer of the codec is never directly accessed by the generated code.

Both experiments were run with Z3 v4.12.2 and cvc5 1.1.2 in the portfolio, with timeouts of 3 minutes and 15 minutes, respectively.

Regarding column \textit{PUS-C services}, services 8 and 18 have two invalid VCs related to preconditions.
They both concern the encoding of \lstinline{IA5String}: the recursive function in charge of encoding the string requires the size of the string (represented as a \lstinline{Vector[Byte]}) to have a certain size dictated by the ASN.1 definition.
The caller may not satisfy this condition even in the presence of a constraint check automatically inserted by ASN1SCC.
The latter only validates the position of a null-terminator without knowing the pointer or array's underlying capacity.
Integrating this check is surprisingly non-trivial since it would impact the C and Ada backends.
In particular, the generated C code uses raw pointers without embedding any capacity or size information.
Finally, the two timeouts, both appearing in service 15, are of the same nature, with the difference being that Stainless cannot find a counterexample.

Though it does not stand out from \textit{\lstinline{TC-Packet}}, we note that this \lstinline{SEQUENCE} has many ACN-inserted fields, nested and direct. A prior attempt at proving invertibility for a subset shows that the outlining discussed in \ref{subsec:acn-spec} is necessary.

\subsection{Identified Bugs}
\label{sec:identified-bugs}
We give an overview of some issues we have found thanks to verification.

\textit{\textbf{Incorrect treatment of \lstinline{NaN}.}}
While translating the floating point encoding and decoding functions from C to Scala, we have discovered that an assertion did not hold when the bit pattern represented a \lstinline{NaN}. The original C and Ada code did not handle this case, and we have opened an issue that was swiftly addressed\footnote{\url{https://github.com/maxime-esa/asn1scc/issues/287}}. Note that we represent the floating point number as an uninterpreted \lstinline{Long} since Stainless does not support \lstinline{Double} or \lstinline{Float}.

\textit{\textbf{Improper alignment for padding.}}
\lstinline{SEQUENCE} whose fields have an alignment requirement were not correctly aligned with padding bits. Additionally, the buffer was not appropriately checked for remaining space in such cases\footnote{\url{https://github.com/maxime-esa/asn1scc/issues/283}}.

\textit{\textbf{Erroneous decoding for optional \lstinline{CHOICE}.}}
When prototyping a proof for invertibility, we discovered that optional \lstinline{CHOICE}s were not correctly decoded when the determinant and presence bits were specified with ACN\footnote{\url{https://github.com/maxime-esa/asn1scc/issues/289}}.

\textit{\textbf{Missing validation check for 7-bit strings.}}
For all backends, 7-bit strings are represented as byte arrays.
However, the generated encoder does not validate the bytes' values and assumes they are within the range [0, 127], which leads to incorrectly written data.
\section{Conclusion}
We developed a Scala backend for the ASN1SCC compiler with an accompanying library for ASN.1/ACN primitives. We proved the absence of runtime errors for the library and the generated code. We also proved that the deserialization functions from the library used by the PUS-C services are the inverse of the corresponding serialization functions. For the generated code, we also established the precise amount of data written or read. Furthermore, we proved the invertibility property for records. Our verification and specification process led us to discover several bugs in the existing code generator. The errors have been addressed, leading to more reliable communication infrastructure.

\section*{Acknowledgements} This research was supported by the European Space Agency Open Space Innovation Platform, 4000140196/22/NL/GLC/ov, New Concepts for Onboard Software Development. We thank Maxime Perrotin for overseeing the project.

\goodbreak
{\raggedcolumns
\bibliographystyle{splncs04}
\bibliography{bib}
}

\end{document}